\documentclass[12pt,preprint]{aastex}

\newcommand{\be}{\begin{equation}}
\newcommand{\ee}{\end{equation}}

\shorttitle{Multi-Color Photometry of $Kepler$ Planet Candidates}
\shortauthors{Col\'on \& Ford}

\begin{document}
\title{Vetting $Kepler$ Planet Candidates with Multi-Color Photometry from the GTC:  Identification of an Eclipsing Binary Star Near KOI 565}

\author{Knicole D. Col\'on\altaffilmark{1,2}, Eric B.\ Ford\altaffilmark{1}}

\altaffiltext{1}{Department of Astronomy, University of Florida, Gainesville, FL 32611-2055, USA.}
\altaffiltext{2}{NSF Graduate Research Fellow.}

\begin{abstract} 
We report the discovery of an eclipsing binary star (KIC 7025851) near KOI 565 (KIC 7025846) based on photometric observations of KOI 565 and several nearby stars acquired in two narrow bandpasses (790.2 and 794.3$\pm$2.0 nm) nearly-simultaneously with the GTC/OSIRIS.  We use the individual photometry in each bandpass as well as the colors of KOI 565 and other nearby stars to determine that the source of the transit signal initially detected by $Kepler$ is not due to a super-Earth-size planet around KOI 565.  Instead, we find the source to be a background eclipsing binary star located $\sim$15 arcsec to the North of KOI 565.  We discuss future prospects for using high-precision multi-color photometry from the GTC to determine whether additional $Kepler$ planet candidates have a planetary nature or are instead false positives (e.g., foreground or background eclipsing binaries or hierarchical triple systems).  Our approach is complementary to multi-color follow-up observations of $Kepler$ planet candidates currently being conducted with the $Spitzer$ space telescope in the infrared as well as to other follow-up techniques.
\end{abstract}

\section{Introduction} 
\label{intro}

At present, there are over 180 confirmed transiting planets, but only $\sim$10\% are estimated to be Neptune-size or smaller.\footnote{The Extrasolar Planets Encyclopedia; http://exoplanet.eu/}  The $Kepler$ space mission, which launched in 2009, is responsible for the discovery of a majority of the known transiting Neptune- and super-Earth-size planets.  Some of the small planets $Kepler$ has discovered to date include Kepler-4b \citep[a Neptune-size planet;][]{borucki10}, Kepler-9d \citep[a super-Earth-size planet in a system with two Saturn-size planets;][]{holman10,torres11}, Kepler-10b \citep[a rocky planet;][]{batalha11}, and Kepler 11-b, c, d, e, f, g \citep[6 Neptune- to super-Earth-size planets;][]{lissauer11}.  Further, $Kepler$ recently discovered over 1000 additional small planet candidates \citep[$R_p$ $<$ 6$ R_{\oplus}$;][]{borucki11b}.  While 80-95\% of these candidates are expected to be true planets \citep{borucki11b,morton11}, identifying which are false positives remains a challenge.
  
The main sources of false positives are background (or, rarely, foreground) eclipsing binaries (EBs) or hierarchical multiple systems.  Due to $Kepler$'s large PSF ($\sim$6 arcsec), the flux from a star that has an eclipsing stellar or planetary companion can be blended with the flux of $Kepler$'s target star if the two stars are spatially co-aligned with each other.  In these cases, it appears that the target star has a transiting companion.  High-resolution ground-based imaging (adaptive optics or speckle imaging), spectroscopy, and $Kepler$'s centroid analysis help to eliminate many blends, but can struggle in cases where the blended system is separated by less than $\sim$0.1 arcsec from the target \citep[e.g.,][]{borucki11b}.  Further, radial velocity (RV) follow-up is very time-consuming for the typical $Kepler$ target, which may be fainter than $V$$\sim$14 \citep{batalha10}.  In the absence of RV measurements, the detection of secondary transits or differences in the depths of individual transits can also be used to help rule out a blended system.  Here, we consider an alternative technique that (to the best of our knowledge) was first discussed by \citet{tingley04} and first demonstrated observationally by \citet{odonovan06}, which is to rule out blends by measuring the transit depth in different bandpasses.  This is possible because, as shown by \citet{tingley04}, the color change during a transit event (i.e. the difference in the transit depth measured at different wavelengths) increases as the color between the different components of a blend increases.  Therefore, observations acquired in multiple bandpasses can be used to reject a planet candidate if the measured transit depths in different bandpasses differ significantly, which could indicate, for example, a blend with a stellar EB of a different spectral type than the target star \citep{tingley04,odonovan06,torres11}.  We note that the $CoRoT$ space telescope even has a prism built-in for the purpose of vetting $CoRoT$ planet candidates with multi-color photometry, further demonstrating the value of such a technique \citep[][]{auvergne09,deeg09}.

Here, we describe multi-wavelength observations acquired with the Optical System for Imaging and low Resolution Integrated Spectroscopy (OSIRIS) installed on the 10.4-m Gran Telescopio Canarias (GTC) that we used to determine the true nature of ($Kepler$ Object of Interest) KOI 565.01.\footnote{Also known as KIC 7025846 in the $Kepler$ Input Catalog.}  KOI 565.01 ($Kepler$ mag = 14.3) was presented by \citet{borucki11a} as a super-Earth-size planet candidate, with an estimated planet radius of $\sim$1.6 $R_{\oplus}$, orbiting a 1.068 $R_{\odot}$ star with a period of 2.34 days.  However, the recent paper by \citet{borucki11b} lists KOI 565.01 as most likely being a false positive, as $Kepler$ measured a centroid shift of $\sim$8 arcsec to the North of KOI 565, indicating that the star that is actually dimming is located on a pixel that is offset from the position of KOI 565.  Given that we did not have this information at the time of the observations presented here, we operate under the assumption that we did not know whether KOI 565.01 was a true planet or false positive.  In this case, it was possible for us to resolve the true source of the transit signal that contaminated $Kepler$'s photometry of KOI 565.  We also present measurements of the color of KOI 565 and several nearby stars during the predicted transit event, which independently confirm that KOI 565 is in fact not the true host of the transit signal.  More importantly, we show that the false positive would have been identified even if the separation between the stars was too small to either spatially resolve them or allow for the measurement of a centroid shift.  Our approach offers an efficient false-positive identification method that is highly complementary to the multi-color follow-up photometry that is currently being conducted with the $Spitzer$ space telescope at infrared wavelengths \citep[e.g.,][]{fressin11} as well as to other ground-based follow-up techniques.

We describe our observations in \S\ref{obs} and the data reduction and light curve (LC) analysis in \S\ref{reduction}.  In \S\ref{results} we present our results and demonstrate that color photometry from the GTC can be used to help identify false positives from transit surveys.  Finally, in \S\ref{discuss}, we conclude with a summary of our results and a discussion of our plans for future observations of additional $Kepler$ planet candidates with the GTC.

\section{Observations}
\label{obs}

We observed the target and several nearby stars around the predicted time of the transit event on 2010 September 19 using the OSIRIS tunable filter (TF) imager installed on the 10.4-m GTC.  With the TF imager, custom bandpasses with a central wavelength between 651-934.5 nm and a FWHM of 1.2-2.0 nm can be specified.  In this observing mode, the effective wavelength decreases radially outward from the optical centre, so we positioned the target and a ``primary" reference star (i.e., most comparable in brightness to the target) at the same distance from the optical centre so that both stars would be observed at the same wavelengths.  Several ``secondary" reference stars were also observed, but they were all at different distances from the optical centre and thus were observed at slightly different wavelengths.  During the observations, we alternated between two bandpasses centred on 790.2 and 794.3 nm (at the location of the target on the CCD chip) and with FWHM of 2.0-nm.  These bandpasses were specifically chosen as they minimize effects of telluric absorption and emission and yield extremely high differential precisions as demonstrated by \cite{colon2010b}.  We used 1$\times$1 binning, a fast pixel readout rate of 500 kHz, and read out a single window (containing the target and several reference stars) located on one CCD chip of 1415 $\times$ 2830 pixels (equivalent to $\sim$ 3$\times$6 arcmin or $\sim$56\% of the CCD chip) in order to decrease the dead time between exposures.  Due to the faintness of the target ($V$ $\sim$ 14.3) and the narrow bandpasses used, the exposure time (for both filter settings) was set to 180-s, with each exposure followed by approximately 21-s of dead time.  

The observations began at 21:52 UT on 2010 September 19 (during bright time) and ended the following morning at 01:55 UT.  There were thin cirrus clouds around the time of observations.  The airmass ranged from $\sim$1.07 to 2.27.  The actual seeing was better than 1.0 arcsec, but the telescope was intentionally defocused to reduce pixel-to-pixel sensitivity variations, so the defocussed FWHM of the target varied between $\sim$1.3-2.0 arcsec ($\sim$10.2-15.5 pixels).  The telescope's guiding system kept the images aligned within a few pixels during the observations, with the target's centroid coordinates shifting by $<$ 2 pixels in either direction.

The predicted mid-transit time based on the ephemeris and orbital period from \citet{borucki11a} was 23:59 UT (2455459.502 BJD) on 2010 September 19.  However, \citet{borucki11b} presented an updated ephemeris and orbital period, so the transit event we observed occurred $\sim$135 min later than initially predicted; because of this, our observations ended before mid-transit.  It should be noted that the updated ephemeris from \citet{borucki11b} is still consistent with the uncertainty in the original ephemeris from \citet{borucki11a}.  Furthermore, the ephemerides given in \citet{borucki11a} were based on $\sim$43 days of observations, while \citet{borucki11b} cited ephemerides based on a much longer time baseline, thus allowing for significantly more precise constraints on the ephemerides.  

\section{Data Reduction and Analysis}
\label{reduction}

Standard IRAF procedures for bias subtraction and flat-field correction were used.  In total, 95 dome flats were taken for each filter setting.  We note that the dome lights do not produce a uniform illumination, so we added an illumination correction to the final flat-field image.  Due to the narrow filters used and position-dependent wavelength, all images contain sky (OH) emission rings.  Therefore, we performed sky subtraction on all images using the IRAF package TFred,\footnote{Written by D. H. Jones for the Taurus Tunable Filter, previously installed on the Anglo-Australian Telescope; http://www.aao.gov.au/local/www/jbh/ttf/adv\_reduc.html} which measures the sky background while including the rings due to sky emission.  We then performed aperture photometry on the target and several reference stars using standard IDL routines.  We tested several different size apertures and based our final choice of aperture on that which yielded the smallest scatter in the relative flux ratios (i.e., the target flux divided by the total reference flux) outside of transit.  The final aperture radius used in our analysis is 23 pixels ($\sim$2.9 arcsec).  No sky annulus was needed due to the use of TFred, which automatically removes the sky background.  These procedures were performed for each filter separately, but we considered the results for each bandpass and used the same aperture for each filter.  We discarded 3-4 images taken in each bandpass due to errors in the reduction process that prohibited us from performing aperture photometry on these images.

LCs for each bandpass were computed for KOI 565 by dividing the flux measured within the target aperture by the total weighted flux of an ensemble of reference stars.  We used six reference stars in total to compute the reference ensemble flux.  Even though these reference stars were located at different distances from the optical centre than the target, we found that using an ensemble of reference stars rather than just a single reference star greatly improved the signal-to-noise (S/N) ratio of our observations.  Each LC was then normalized to the mean baseline (out-of-transit; OOT) flux ratio as measured in each bandpass.  We corrected each LC against changes in the airmass, and in order to account for any additional systematics in the LCs, we performed external parameter decorrelation \citep[EPD; see, e.g.,][]{bakos07,bakos10} against each of the following parameters: the X and Y centroid coordinates of the target on the image frames and the sharpness of the target's profile [equivalent to (2.35/FWHM)$^2$].

The photometric uncertainties in the relative flux ratios include the photon noise of the target and the reference ensemble, the noise in the sky background around the target and references, and scintillation noise.  We calculate the median photometric uncertainties to be 0.987 mmag for both the 790.2 and 794.3-nm LCs, where the photon noise of the target dominates the errors.  We also investigated the possibility of red noise within our data by computing the standard deviation of the flux ratios after averaging the data over several different bin sizes, and we found that our data follows the trend expected for white Gaussian noise. 

\section{Results}
\label{results}

\begin{figure}
\plotone{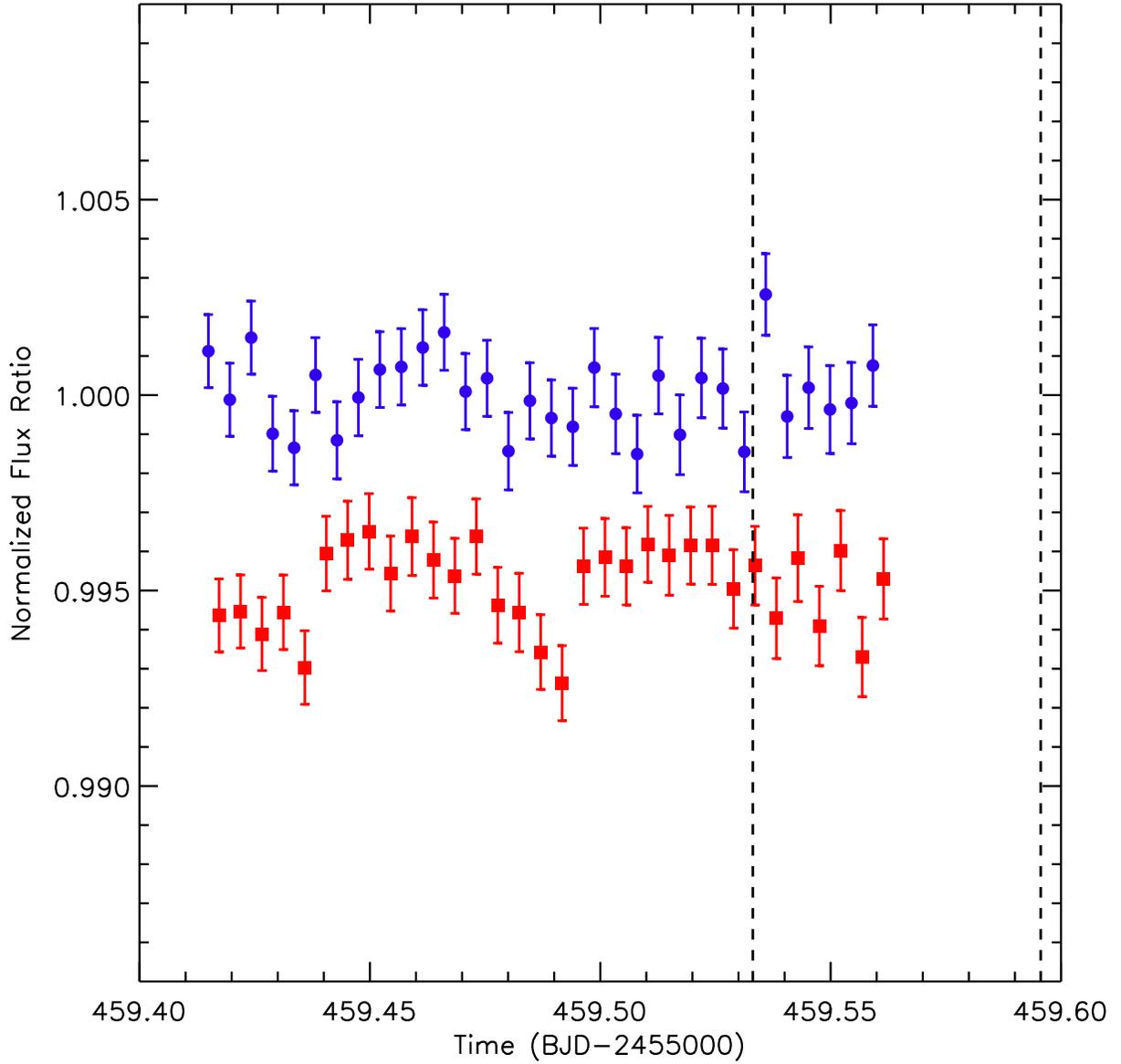}
\caption{Normalized LCs for nearly-simultaneous observations at 790.2$\pm$2.0 nm (blue circles) and 794.3$\pm$2.0 nm (red squares) of KOI 565 as observed on 2010 September 19.  The 794.3-nm LC has been offset for clarity.  The vertical dashed lines indicate (from left to right) the predicted beginning of ingress and mid-transit time [based on \citet{borucki11b}].
\label{koilc}}
\end{figure}

We present the results of our photometry in Table \ref{koitab} and the corresponding LCs for KOI 565 in Figure \ref{koilc}.\footnote{We note the observation times given are the Barycentric Julian Dates in Barycentric Dynamical Time (BJD\_TDB), computed from the Julian Dates using the calculator found at http://astroutils.astronomy.ohio-state.edu/time/utc2bjd.html \citep{eastman10}.  The ephemerides given by $Kepler$ are in the same time coordinate system.}  Based on the parameters of the candidate planet transiting KOI 565 given by \citet{borucki11a}, we expected to measure a transit depth of approximately 182 ppm.  Due to the somewhat low S/N ratio of our observations (the result of using very narrow bandpasses to observe a faint target while maintaining a reasonable exposure time), our photometric precisions (0.978 mmag) were insufficient to detect such a shallow transit.  However, we did not need very high precisions to determine that another star near the target was the true source of the dimming that $Kepler$ observed in KOI 565.  

\begin{figure}
\plotone{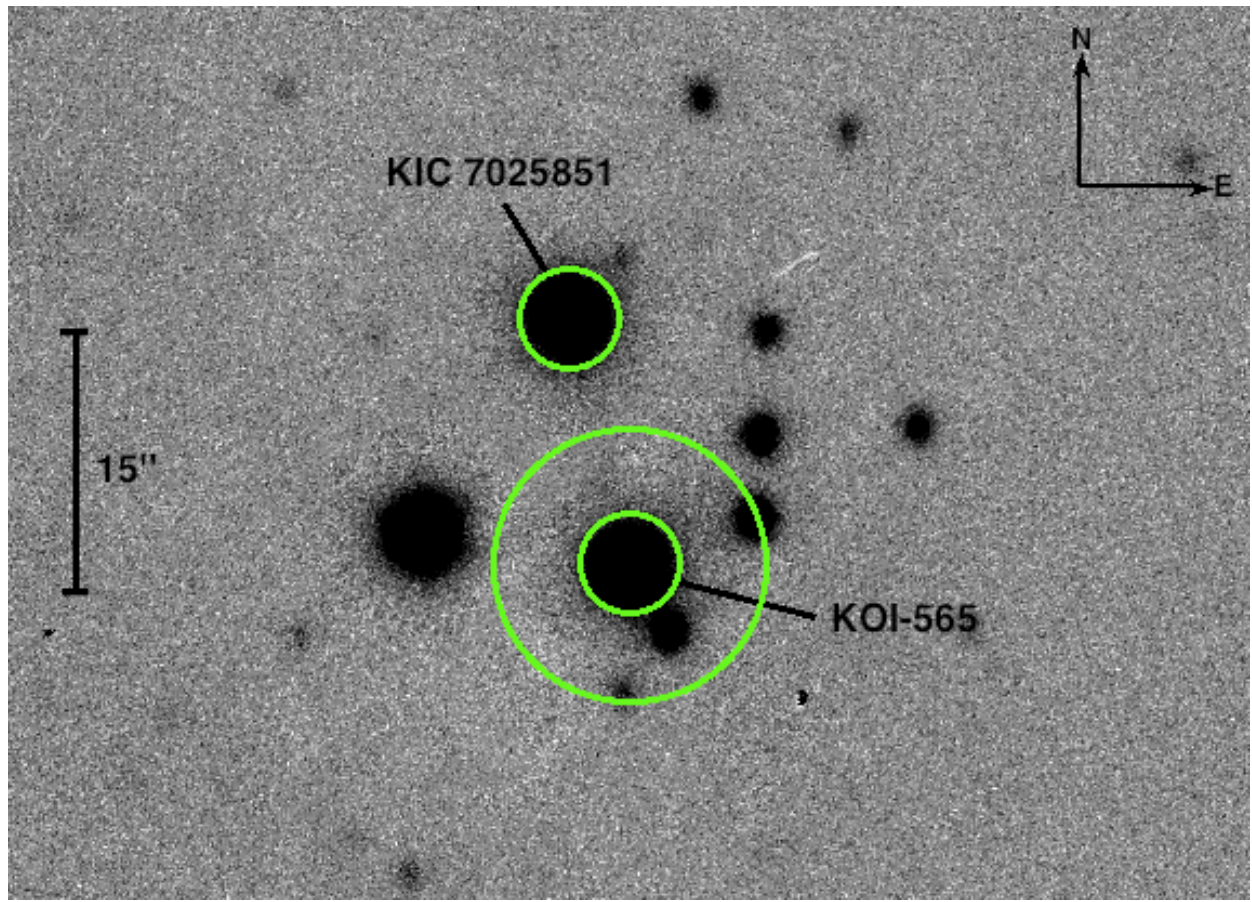}
\caption{Image from GTC/OSIRIS observations at 790.2-nm containing part of the field of view around KOI 565.  This is only a small portion of the observed field of view, so the six reference stars used in our analysis are not shown here.  The target is located at $\alpha$ = 19$^{h}$17$^{m}$26.05$^{s}$, $\delta$ = 42$^{\circ}$31$^{'}$34.3$^{''}$, and KIC 7025851 is located at $\alpha$ = 19$^{h}$17$^{m}$26.30$^{s}$, $\delta$ = 42$^{\circ}$31$^{'}$48.9$^{''}$.  The small green circles around KOI 565 and KIC 7025851 indicate the size of the apertures used in our photometry ($r$ $\sim$ 23 pixels $\sim$ 2.9 arcsec).  The larger green circle around KOI 565 has a radius of 8 arcsec and is included simply to illustrate which stars are located within a distance 8 arcsec from the target, as \citet{borucki11b} measured a centroid shift of 8 arcsec to the North of the target.  The position of the star causing the centroid shift is typically at a slightly further distance, so the position of KIC 7025851 relative to KOI 565 is consistent with measurements from $Kepler$.
\label{fov}}
\end{figure}

\begin{figure}
\plotone{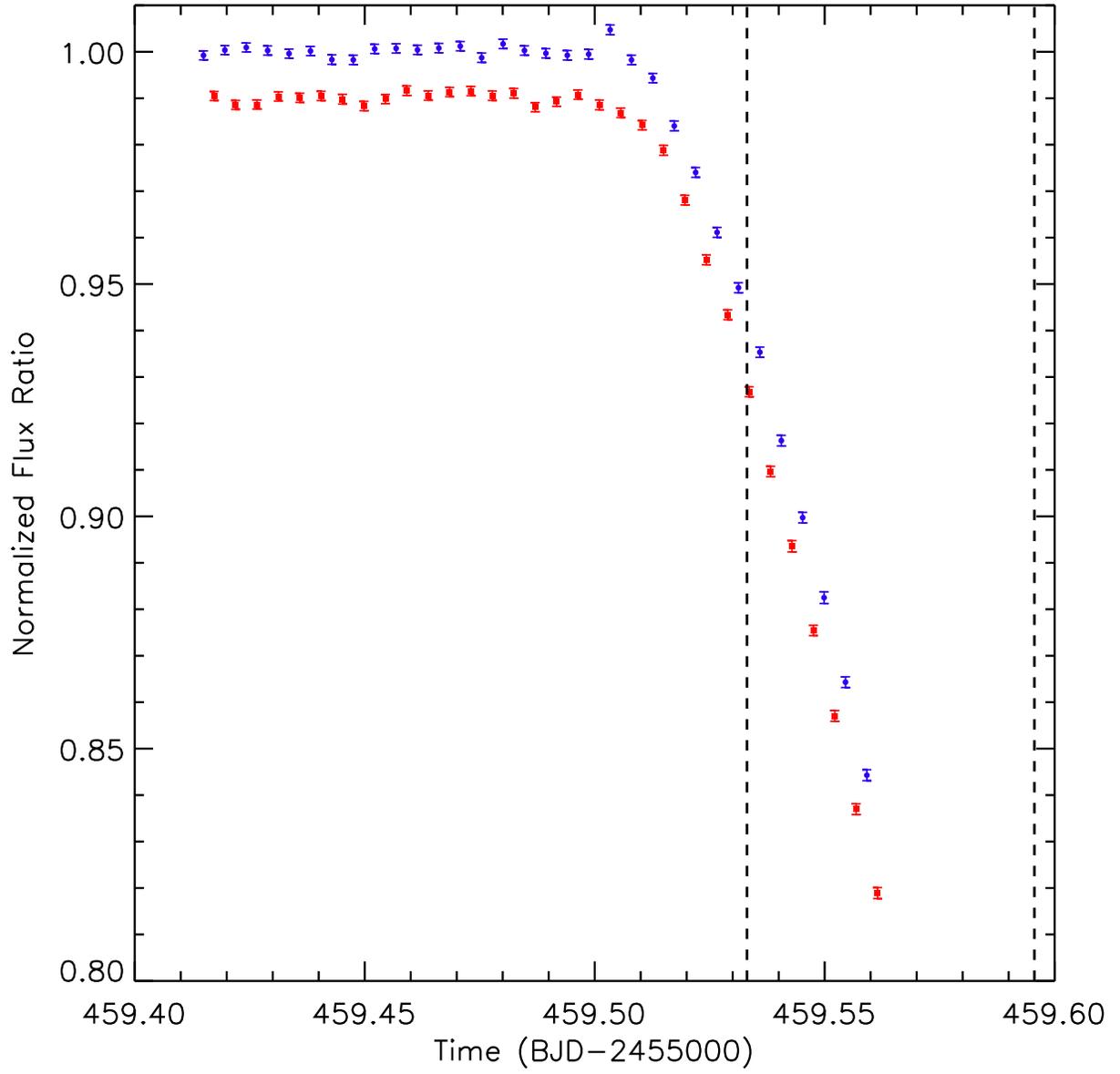}
\caption{Similar to Figure \ref{koilc}, but for KIC 7025851.  While only a partial eclipse was observed, the minimum depth of the LC is comparable to what is expected for a stellar EB.
\label{sourcelc}}
\end{figure}

We followed similar procedures as described in \S\ref{reduction} to compute the LCs of several stars within $\sim$20 arcsec of the location of the target that might have contaminated $Kepler$ photometry of KOI 565, and we visually inspected their LCs to see if any showed a transit-like signal around the predicted time of the transit.  For reference, we present the field of view around KOI 565 in Figure \ref{fov}.  From our analysis of stars near KOI 565, we determined that a star (KIC 7025851) approximately 15 arcsec to the North of KOI 565 is the true source of the transit signal, as we observed a significant decrease ($>$ 15\%) in the brightness of that star at the time of the predicted transit event.  We present the LCs for KIC 7025851 in Figure \ref{sourcelc}, and the photometry is also given in Table \ref{sourcetab}.  Although we were not able to observe a complete LC, based on both the minimum depth and shape of the LC, we deduce that KIC 7025851 is a stellar EB.  Thus, our identification of KIC 7025851 as a stellar EB that contaminated $Kepler$'s photometry of KOI 565 is consistent with the magnitude and direction of the centroid shift as well as the eclipse ephemeris from the $Kepler$ data \citep{borucki11b}.  The EB therefore has an eclipse ephemeris of 2455459.5956 BJD and orbital period of 2.340506 days [as determined by \citet{borucki11b}].  Based on the stars colors and relative brightnesses, we infer that KIC 7025851 is a background (rather than foreground) EB.

\begin{figure}
\epsscale{0.75}
\plotone{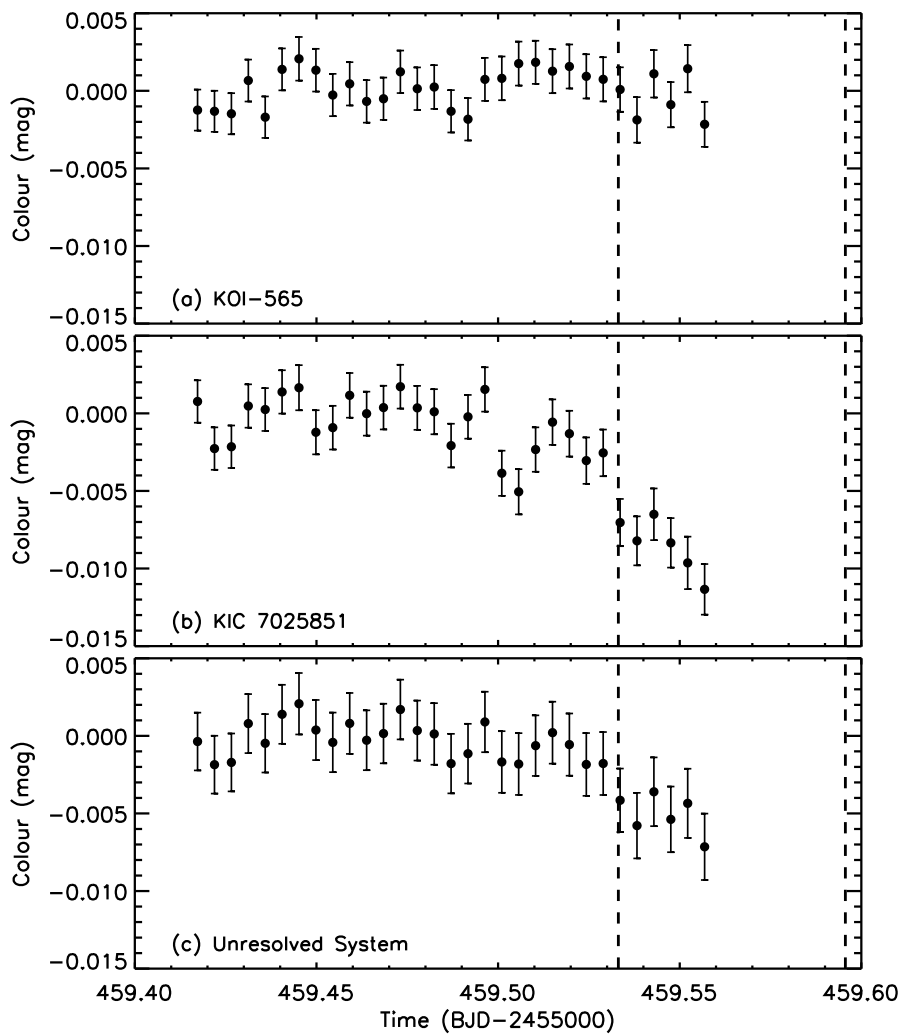}
\caption{The colors as computed between the 790.2 and 794.3 nm observations of (a) KOI 565, (b) a stellar EB (KIC 7025851), and (c) for an ``unresolved" system (the target light combined with the light from the EB).  In each panel, the vertical dashed lines indicate (from left to right) the approximate beginning of ingress and the mid-transit time \citep[based on][]{borucki11b}.  The vertical scale is the same for each panel for ease of comparison.  There is no change in the color seen for KOI 565, but for the EB as well as the hypothetical unresolved system, we measure an appreciable difference in the color during the transit event.
\label{colors}}
\end{figure}

Next, we consider the colors (790.2-nm $-$ 794.3-nm) of the stars during the transit event.  In Figure \ref{colors}, we present the colors of KOI 565 and KIC 7025851.  We also consider the color for an ``unresolved" system, simulating a scenario in which the target star and EB are physically associated and thus could not be spatially resolved so that all their light is combined.  For this case, we combine the flux from KOI 565 with the flux from KIC 7025851 in each bandpass and then compute the color from the LCs of the unresolved system.\footnote{Based on the separation of the target and the EB, it is most likely that only a portion of the light from the EB was blended with the light from KOI 565.  See Figure \ref{fov} for further reference.}  Letting $\lambda_{1}$ be the apparent magnitude in the 790.2-nm bandpass and $\lambda_{2}$ be the magnitude in the 794.3-nm bandpass, we calculated the color indices as
\be
\lambda_{1} - \lambda_{2}=-2.5\log{\frac{F_{\lambda_{1}}}{F_{\lambda_{2}}}},
\ee
where we have taken the average of each pair of flux ratios in the 790.2-nm LC ($F_{\lambda_{1}}$) and divided by the corresponding points in the 794.3-nm LC ($F_{\lambda_{2}}$).  As illustrated in Figure \ref{colors}, we do not measure an appreciable change in the color of KOI 565 during the transit event, but the color of KIC 7025851 shows a significant color change.  To compare the color changes for each star directly, we calculate the weighted mean colors and their uncertainties for the interval before the transit (i.e., the interval to the left of the leftmost dashed line in Figure \ref{colors}) and compare those to the same values measured during the partial transit event.  For KOI 565, we calculate the difference in the mean colors to be $6.64\pm6.62\times10^{-4}$, which is consistent with there being no difference in the colors at a level of 1$\sigma$.  On the contrary, the mean colors of KIC 7025851 differ at a confidence level of $\sim$8.3$\sigma$, with a difference of $77.9\pm7.2\times10^{-4}$.  In the case of the hypothetical unresolved system, where we imagine the target and EB to be physically bound so the projected separation between the two stars is undetectable and all their light is added together, we would still measure an appreciable change in the color, with a difference of $47.9\pm9.6\times10^{-4}$ at a significance of $\sim$3.8$\sigma$.  The fact that we measure this large of a color change over such a narrow wavelength regime ($\sim$4-nm) during the transit event clearly indicates a non-planetary source of the color change, i.e., a stellar EB composed of two stars with very different temperatures.  For comparison, \citet{colon2010b} found no appreciable difference between the in-transit and out-of-transit colors as measured from the same bandpasses used here for either TrES-2 or TrES-3, both of which host Jupiter-size planets.

\section{Discussion}
\label{discuss}

In this paper we have presented observations of a $Kepler$ planet candidate acquired in two very narrow bandpasses.   From our observations, we identified a nearby stellar EB in eclipse at the predicted time of the transit of the $Kepler$ candidate.  We also used our observations to measure the change in the color of a hypothetical unresolved source (composed of the $Kepler$ target and the stellar EB) during the transit event.  The identification of a nearby stellar EB and the measured color change during the transit event  separately identify the $Kepler$ candidate as a false positive, thus confirming the findings of \citet{borucki11b}.  Based on the LCs of the resolved stars, we deduce that some of the light from a background stellar EB (KIC 7025851) contaminated the photometry of KOI 565 to mimic the transit of a super-Earth-size planet around KOI 565.

The technique we describe in this paper is complementary to other follow-up observations of transiting planet candidates that are currently being conducted.  For example, the $Spitzer$ space telescope is also being used for follow-up of $Kepler$ planet candidates with a wide infrared bandpass \citep[e.g.,][]{fressin11}, and the $CoRoT$ space telescope has a prism built-in for the purpose of detecting changes in the color during transits of $CoRoT$ planet candidates via three wide optical channels \citep{auvergne09,deeg09}.  However, while $Spitzer$ and $CoRoT$ will go offline in the near-future, our approach of acquiring ground-based transit photometry nearly-simultaneously in narrow optical bandpasses can be used indefinitely.  When compared to multi-color photometry acquired with other ground-based telescopes, our technique has the advantage of being able to acquire multi-color photometry in a single transit observation.  We note that some other ground-based instruments are capable of similar observations, e.g., the Simultaneous Quad IR Imager \citep[SQIID;][]{ellis93} at Kitt Peak National Observatory and ULTRACAM \citep{dhillon07} at the William Herschel Telescope.  However, the GTC/OSIRIS has a unique combination of a large field of view, a superior collecting area, and a wide selection of filters, which combined allows for very efficient high-precision multi-color photometry of faint $Kepler$ targets.

Multi-color photometry with the GTC is thus a useful tool for identifying false positives in transit surveys, since the magnitude of the color change during transit can be used to identify not only background (or foreground) EB stars but also physical triple star systems, which are difficult to reject as they dilute the transit depth and result in a negligible centroid shift.  In the case of physical triples that are difficult to resolve spatially (e.g., with separations $<$ 0.1 arcsec), high-precision multi-color photometry can be useful, as a measurable color change during transit could indicate a blend with a stellar EB of a different spectral type than the target star \citep{tingley04,odonovan06,torres11}.  \citet{morton11} predict a slightly higher false positive rate for $Kepler$ planet candidates due to physical triples than background EBs.  This is in part because physical triples can often mimic the transit depth of Neptune-size planet candidates \citep{morton11}, and Neptune-size planet candidates dominate the candidates discovered by $Kepler$ \citep{borucki11b}.   Thus, high-precision multi-color photometry may be particularly useful for rejecting false positives within the class of Neptune-size planet candidates.  

A blend with a star that is hosting a transiting Jupiter- or Neptune-size planet will be more difficult to reject with multi-color photometry, as the magnitude of the color change during transit will be much smaller than for a blend with a stellar EB.  The most difficult scenario to reject with multi-color photometry is a hierarchical triple system, where a physically associated companion star has a planetary companion.  Our measurements presented in this paper would not have been sensitive to a blend with a star hosting a Jupiter-size planet.  However, we emphasize that our results were based on observations in two narrow bandpasses with central wavelengths that differed by only $\sim$4-nm.  Future observations similar to those here could be conducted using OSIRIS's broadband filters.  These broadband filters are much wider than those allowed by the TF imaging mode, but are still narrower than typical Sloan filters by a factor of $\sim$2-4 so they still reduce effects of differential extinction and variable atmospheric absorption.  The advantage of using slightly wider filters is to ensure that high S/N ratios are achieved even for fainter targets ($V\sim$ 14-15) while maintaining a reasonable exposure time ($<$ a few minutes).  Further, a larger wavelength regime can be covered by observing in, for example, a bluer filter (e.g., $\sim$666-nm) and a redder filter (e.g., $\sim$858-nm), which can enhance the change in the color during a transit event.  Observing in a red filter in particular will also help reduce stellar limb darkening \citep[compared to the broad optical bandpass used by $Kepler$;][]{colon2009}, so that for candidates determined to be true planets, measuring the transit depth in a red bandpass will improve estimates of the planet radius, density and thus bulk composition \citep[assuming a certain mass range for the candidate planet;][]{valencia07}.  While TF imaging is particularly well-suited for high-precision transit photometry of brighter targets \citep[see, e.g., ][]{colon2010a,colon2010b}, it is not ideal for fainter stars due to the longer exposure times required to get a high S/N ratio.  The use of OSIRIS's broader filters that cover a larger wavelength regime is thus one possible way to boost the S/N ratio to reject blends with background, foreground or physically associated stars hosting transiting planets.  

With transit surveys like $Kepler$ and $CoRoT$ actively searching for and finding new planet candidates, it will be vital to use all the tools at hand to reject false positives and determine the true nature of the candidate planets.  These observations demonstrate that multi-color photometry from the GTC is one additional tool that can help with the identification of false positives in the coming years.

\acknowledgements We gratefully acknowledge the observing staff at the GTC and give a special thanks to Ren\'e Rutten, Antonio Cabrera Lavers and Riccardo Scarpa for helping us plan and conduct these observations successfully.  We are extremely grateful to David R. Ciardi, Steve Howell, Suvrath Mahadevan, Avi Shporer and Martin Still for their helpful comments.  We thank the referee for helping us to improve this manuscript.  This material is based upon work supported by the National Science Foundation Graduate Research Fellowship under Grant No. DGE-0802270.  This work is based on observations made with the Gran Telescopio Canarias (GTC), installed in the Spanish Observatorio del Roque de los Muchachos of the Instituto de Astrof\'isica de Canarias, on the island of La Palma.  The GTC is a joint initiative of Spain (led by the Instituto de Astrof\'isica de Canarias), the University of Florida and Mexico, including the Instituto de Astronom\'ia de la Universidad Nacional Aut\'onoma de M\'exico (IA-UNAM) and Instituto Nacional de Astrof\'isica, \'Optica y Electr\'onica (INAOE).

\clearpage

\begin{deluxetable}{cccc}
\tabletypesize{\small}
\tablewidth{0pt}
\tablecaption{Normalized Photometry of KOI 565 \label{koitab}}
\tablehead{
$\lambda$ (nm) & BJD-2455000 & Flux Ratio & Uncertainty}
\startdata
790.2 &     459.41494     &     1.00112     &     0.00094     \\
790.2 &     459.41959     &     0.99988     &     0.00093     \\
790.2 &     459.42424     &     1.00147     &     0.00094     \\
790.2 &     459.42890     &     0.99901     &     0.00096     \\
790.2 &     459.43355     &     0.99865     &     0.00095     \\
790.2 &     459.43821     &     1.00051     &     0.00096     \\
790.2 &     459.44286     &     0.99884     &     0.00099     \\
790.2 &     459.44751     &     0.99994     &     0.00098     \\
790.2 &     459.45216     &     1.00065     &     0.00097     \\
790.2 &     459.45682     &     1.00072     &     0.00098     \\
790.2 &     459.46147     &     1.00122     &     0.00097     \\
790.2 &     459.46612     &     1.00161     &     0.00097     \\
790.2 &     459.47077     &     1.00009     &     0.00097     \\
790.2 &     459.47543     &     1.00043     &     0.00097     \\
790.2 &     459.48008     &     0.99857     &     0.00099     \\
790.2 &     459.48473     &     0.99985     &     0.00097     \\
790.2 &     459.48939     &     0.99941     &     0.00098     \\
790.2 &     459.49404     &     0.99919     &     0.00099     \\
790.2 &     459.49869     &     1.00070     &     0.00100     \\
790.2 &     459.50335     &     0.99952     &     0.00102     \\
790.2 &     459.50800     &     0.99849     &     0.00099     \\
790.2 &     459.51265     &     1.00050     &     0.00098     \\
790.2 &     459.51730     &     0.99898     &     0.00102     \\
790.2 &     459.52196     &     1.00044     &     0.00102     \\
790.2 &     459.52661     &     1.00017     &     0.00101     \\
790.2 &     459.53126     &     0.99855     &     0.00102     \\
790.2 &     459.53591     &     1.00257     &     0.00104     \\
790.2 &     459.54057     &     0.99945     &     0.00105     \\
790.2 &     459.54522     &     1.00019     &     0.00104     \\
790.2 &     459.54987     &     0.99963     &     0.00112     \\
790.2 &     459.55453     &     0.99980     &     0.00104     \\
790.2 &     459.55918     &     1.00076     &     0.00104     \\
\multicolumn{4}{c}{} \\
794.3 &     459.41726     &     0.99936     &     0.00093     \\
794.3 &     459.42192     &     0.99946     &     0.00094     \\
794.3 &     459.42657     &     0.99888     &     0.00094     \\
794.3 &     459.43123     &     0.99944     &     0.00095     \\
794.3 &     459.43588     &     0.99802     &     0.00094     \\
794.3 &     459.44053     &     1.00095     &     0.00096     \\
794.3 &     459.44518     &     1.00129     &     0.00100     \\
794.3 &     459.44984     &     1.00152     &     0.00097     \\
794.3 &     459.45449     &     1.00044     &     0.00096     \\
794.3 &     459.45914     &     1.00139     &     0.00100     \\
794.3 &     459.46380     &     1.00078     &     0.00097     \\
794.3 &     459.46845     &     1.00038     &     0.00096     \\
794.3 &     459.47310     &     1.00139     &     0.00096     \\
794.3 &     459.47775     &     0.99962     &     0.00097     \\
794.3 &     459.48241     &     0.99943     &     0.00100     \\
794.3 &     459.48706     &     0.99842     &     0.00096     \\
794.3 &     459.49171     &     0.99762     &     0.00096     \\
794.3 &     459.49637     &     1.00063     &     0.00097     \\
794.3 &     459.50102     &     1.00085     &     0.00099     \\
794.3 &     459.50567     &     1.00062     &     0.00099     \\
794.3 &     459.51032     &     1.00119     &     0.00097     \\
794.3 &     459.51498     &     1.00091     &     0.00102     \\
794.3 &     459.51963     &     1.00116     &     0.00099     \\
794.3 &     459.52428     &     1.00116     &     0.00100     \\
794.3 &     459.52893     &     1.00004     &     0.00100     \\
794.3 &     459.53359     &     1.00064     &     0.00101     \\
794.3 &     459.53824     &     0.99929     &     0.00103     \\
794.3 &     459.54289     &     1.00083     &     0.00111     \\
794.3 &     459.54755     &     0.99909     &     0.00101     \\
794.3 &     459.55220     &     1.00103     &     0.00102     \\
794.3 &     459.55685     &     0.99829     &     0.00101     \\
794.3 &     459.56150     &     1.00030     &     0.00103     \\
\enddata
\tablenotetext{a}{The time stamps included here are the Barycentric Julian Dates in Barycentric Dynamical Time (BJD\_TDB) at mid-exposure.  The flux ratios included here are those that have been corrected using EPD and normalized to the baseline (OOT) flux ratios (see \S\ref{reduction} for further details).}
\end{deluxetable}
\clearpage

\begin{deluxetable}{cccc}
\tabletypesize{\small}
\tablewidth{0pt}
\tablecaption{Normalized Photometry of KIC 7025851 \label{sourcetab} }
\tablehead{
$\lambda$ (nm) & BJD-2455000 & Flux Ratio & Uncertainty}
\startdata
790.2 &     459.41494     &     0.99923     &     0.00097     \\
790.2 &     459.41959     &     1.00036     &     0.00096     \\
790.2 &     459.42424     &     1.00093     &     0.00097     \\
790.2 &     459.42890     &     1.00028     &     0.00099     \\
790.2 &     459.43355     &     0.99961     &     0.00098     \\
790.2 &     459.43821     &     1.00016     &     0.00098     \\
790.2 &     459.44286     &     0.99832     &     0.00102     \\
790.2 &     459.44751     &     0.99826     &     0.00101     \\
790.2 &     459.45216     &     1.00062     &     0.00100     \\
790.2 &     459.45682     &     1.00073     &     0.00100     \\
790.2 &     459.46147     &     1.00041     &     0.00099     \\
790.2 &     459.46612     &     1.00080     &     0.00100     \\
790.2 &     459.47077     &     1.00121     &     0.00100     \\
790.2 &     459.47543     &     0.99873     &     0.00100     \\
790.2 &     459.48008     &     1.00170     &     0.00102     \\
790.2 &     459.48473     &     1.00028     &     0.00100     \\
790.2 &     459.48939     &     0.99967     &     0.00100     \\
790.2 &     459.49404     &     0.99925     &     0.00101     \\
790.2 &     459.49869     &     0.99948     &     0.00103     \\
790.2 &     459.50335     &     1.00473     &     0.00104     \\
790.2 &     459.50800     &     0.99824     &     0.00102     \\
790.2 &     459.51265     &     0.99434     &     0.00101     \\
790.2 &     459.51730     &     0.98403     &     0.00106     \\
790.2 &     459.52196     &     0.97403     &     0.00106     \\
790.2 &     459.52661     &     0.96113     &     0.00106     \\
790.2 &     459.53126     &     0.94922     &     0.00107     \\
790.2 &     459.53591     &     0.93534     &     0.00111     \\
790.2 &     459.54057     &     0.91630     &     0.00113     \\
790.2 &     459.54522     &     0.89972     &     0.00113     \\
790.2 &     459.54987     &     0.88250     &     0.00124     \\
790.2 &     459.55453     &     0.86431     &     0.00116     \\
790.2 &     459.55918     &     0.84428     &     0.00118     \\ 
\multicolumn{4}{c}{} \\
794.3 &     459.41726     &     1.00050     &     0.00097     \\
794.3 &     459.42192     &     0.99855     &     0.00097     \\
794.3 &     459.42657     &     0.99862     &     0.00097     \\
794.3 &     459.43123     &     1.00038     &     0.00099     \\
794.3 &     459.43588     &     1.00011     &     0.00098     \\
794.3 &     459.44053     &     1.00051     &     0.00099     \\
794.3 &     459.44518     &     0.99981     &     0.00104     \\
794.3 &     459.44984     &     0.99832     &     0.00100     \\
794.3 &     459.45449     &     0.99983     &     0.00099     \\
794.3 &     459.45914     &     1.00164     &     0.00103     \\
794.3 &     459.46380     &     1.00059     &     0.00101     \\
794.3 &     459.46845     &     1.00135     &     0.00099     \\
794.3 &     459.47310     &     1.00155     &     0.00100     \\
794.3 &     459.47775     &     1.00054     &     0.00100     \\
794.3 &     459.48241     &     1.00108     &     0.00104     \\
794.3 &     459.48706     &     0.99806     &     0.00099     \\
794.3 &     459.49171     &     0.99926     &     0.00099     \\
794.3 &     459.49637     &     1.00079     &     0.00101     \\
794.3 &     459.50102     &     0.99855     &     0.00103     \\
794.3 &     459.50567     &     0.99683     &     0.00102     \\
794.3 &     459.51032     &     0.99415     &     0.00101     \\
794.3 &     459.51498     &     0.98867     &     0.00106     \\
794.3 &     459.51963     &     0.97785     &     0.00103     \\
794.3 &     459.52428     &     0.96487     &     0.00105     \\
794.3 &     459.52893     &     0.95294     &     0.00107     \\
794.3 &     459.53359     &     0.93619     &     0.00108     \\
794.3 &     459.53824     &     0.91884     &     0.00112     \\
794.3 &     459.54289     &     0.90259     &     0.00122     \\
794.3 &     459.54755     &     0.88429     &     0.00112     \\
794.3 &     459.55220     &     0.86569     &     0.00115     \\
794.3 &     459.55685     &     0.84542     &     0.00115     \\
794.3 &     459.56150     &     0.82719     &     0.00118     \\
\enddata
\tablenotetext{a}{The columns are the same as in Table \ref{koitab}.}
\end{deluxetable}
\clearpage

\clearpage

\clearpage

\clearpage

\end{document}